\documentclass[12pt]{article}
\usepackage[dvips]{graphicx}
\renewcommand{\baselinestretch}{1.20}
\begin{document}

%
%hep-ph/0610419
\begin{flushright}
OU-HET-569, \, November, 2006 \ \ \\
\end{flushright}
\vspace{0mm}
\begin{center}
\large{On the new COMPASS measurement of the deuteron \\
spin-dependent
structure function $g_1^d$}
\end{center}
\vspace{0mm}
\begin{center}
M.~Wakamatsu\footnote{Email \ : \ wakamatu@phys.sci.osaka-u.ac.jp}
\end{center}
\vspace{-4mm}
\begin{center}
Department of Physics, Faculty of Science, \\
Osaka University, \\
Toyonaka, Osaka 560-0043, JAPAN
\end{center}

%\vspace{4mm}
%PACS numbers : 12.39.Fe, 12.39.Ki, 12.38.Lg, 13.60.Hb, 14.20.Dh

\vspace{6mm}
\begin{center}
\small{{\bf Abstract}}
\end{center}
\vspace{-2mm}
\begin{center}
\begin{minipage}{15.5cm}
\renewcommand{\baselinestretch}{1.0}
\small
Very recently, a new measurement of the deuteron spin-dependent
structure function $g_1^d (x)$ was reported by the COMPASS group.
A main change from the old SMC measurement is a considerable
improvement of the statistical accuracy in the low $x$ region
$0.004 < x < 0.03$. We point out that the new COMPASS data
for $g_1^d (x)$ as well as their QCD fits for $\Delta \Sigma$ and
$\Delta s + \Delta \bar{s}$ are all remarkably close to our
theoretical predictions given several years ago based on the
chiral quark soliton model. 

\normalsize
\end{minipage}
\end{center}
\renewcommand{\baselinestretch}{2.0}

\vspace{8mm}
%\section{Introduction}

If the intrinsic quark spin carries little
of the total nucleon spin, what carries the rest of the nucleon
spin? Quark orbital angular momentum (OAM) $L^Q$? Gluon OAM $L^g$?
Or gluon polarization $\Delta g$? That is a still unsolved
fundamental puzzle of QCD \cite{EMC88}.
Toward the solution of the problem, remarkable progress has been made
for the past few years.
First, the new COMPASS measurement of the quasi-real photoproduction
of high-$p_T$ hadron pairs indicates that $\Delta g$ cannot be very
large at least below $Q^2 \leq 3 \,\mbox{GeV}^2$ \cite{COMPASS06G}.
(The small gluon polarization is also indicated by PHENIX measurement
of neutral pion double longitudinal spin asymmetry in the proton-proton
collisions \cite{PHENIX06} and also by the STAR measurement of the double
longitudinal spin asymmetry in inclusive jet production in polarized
proton-proton collisions \cite{STAR05},\cite{STAR06}.)
There also appeared an interesting paper by Brodsky and
Gardner \cite{BG06}, in which, based on the conjecture on the relation
between the Sivers mechanism \cite{Sivers90} and the quark
and gluon OAM \cite{Burk02},
it was argued that small single-spin asymmetry observed by the COMPASS
collaboration on the deuteron target is an indication of
small gluon OAM. These observations together with
the progress of the physics of generalized parton distribution
functions \cite{Ji98} arose a growing interest on the role of quark OAM
in the nucleon.

The possible importance of quark OAM was pointed out many
years ago based on the chiral soliton picture of the nucleon : 
first within the Skyrme model \cite{BEK88}, second
within the chiral quark soliton model (CQSM) \cite{WY91}.
According the latter paper, the dominance of quark OAM
is inseparably connected with collective motion of quarks
in the rotating hedgehog mean field.
The CQSM predicts at the model energy scale around $600 \,\mbox{MeV}$
that $\Delta \Sigma$ is around 0.35, while $2 \,L_q$ is around 0.65.
The CQSM also reproduces well the spin structure functions for
the proton, the neutron and the deuteron \cite{WK99},\cite{Waka03}.
Very recently, a new measurement of the deuteron spin structure
function $g_1^d (x,Q^2)$ was reported by the COMPASS
group \cite{COMPASS05D}. One should recognize that
the precise measurement of $g_1^d (x,Q^2)$ is of crucial importance,
because, aside from small effects of $s$-quark polarization as well
as the nuclear binding effects etc., it is just proportional to the
isosinglet quark helicity distribution, the integral of which gives
the intrinsic quark-spin contribution to the total nucleon spin.

The purpose of the present paper is to point out that the new COMPASS
data for $g_1^d (x)$ improved in the small-$x$ region
turns out to be quite close to our theoretical
predictions given several years ago based on the
CQSM \cite{WK99},\cite{Waka03}.
We also compare our predictions for the polarized strange quark
distribution $\Delta s(x)$ with the recent
QCD fits at the next-to-leading order (NLO) performed by the
COMPASS group, to find that they are order of magnitude consistent.
We shall also see that the net longitudinal quark polarization
$\Delta \Sigma$ as well as the strange quark polarization
$\Delta s + \Delta \bar{s}$ in the nucleon extracted from the recent
QCD fits by the CAMPASS \cite{COMPASS06D} and HERMES \cite{HERMES06D}
collaborations are not only mutually consistent
but also surprisingly close to the predictions of the CQSM.

In QCD, the longitudinal spin structure functions for the proton
and the neutron are given as
\begin{eqnarray}
 g_1^{p/n} (x) &=& \frac{1}{9} \,
 \left( \,C_{NS} \otimes \left[ \,\pm \,\Delta q_3 + 
 \frac{1}{4} \,\Delta q_8 \,\right] 
 \ + \ C_S \otimes \Delta \Sigma + 
 2 \,N_f \,C_g \otimes \Delta g \right) ,
\end{eqnarray}
where $C_{NS}, C_S, C_g$ are the nonsinglet, singlet and gluon
Wilson coefficients, while the symbol $\otimes$ represents the
convolution with the quark and gluon distribution functions :
\begin{eqnarray}
 \Delta q_3 (x) &\equiv& ( \Delta u(x) + \Delta \bar{u}(x)) \ - \ 
 ( \Delta d(x) + \Delta \bar{d}(x)), \\
 \Delta q_8 (x) &\equiv& ( \Delta u(x) + \Delta \bar{u}(x)) \ + \ 
 ( \Delta d(x) + \Delta \bar{d}(x)) \ - \ 
 2 \,( \Delta s(x) + \Delta \bar{s}(x)), \\
 \Delta \Sigma (x) &\equiv& ( \Delta u(x) + \Delta \bar{u}(x))
 \ + \  ( \Delta d(x) + \Delta \bar{d}(x)) \ + \ 
 ( \Delta s(x) + \Delta \bar{s}(x)).
\end{eqnarray}
It is customary to assume that the structure functions $g_1^p (x)$
and $g_1^n (x)$ on proton and neutron targets are related to that
of the deuteron by the relation
\begin{equation}
 g_1^d (x) \ = \ \frac{1}{2} \,(g_1^p (x) + g_1^n (x)) \,
 \left( \,1 - \frac{3}{2} \,\omega_D \right),
\end{equation}
with $\omega_D$ the $D$-state admixture to the deuteron wave
function. Instead of $g_1^d (x)$, it is then more convenient
to use $g_1^N (x) \equiv g_1^d (x) / (1 - \frac{3}{2} \,\omega_D)$,
in which the correction for the $D$-state admixture in the deuteron
state has been taken into account.

At the leading order (LO), $g_1^p$ and $g_1^n$ reduce to
\begin{eqnarray}
 g_1^p (x) &=& \frac{1}{9} \,\left[ \,4 \,\Delta u(x) + 
 \Delta d(x) - \Delta s(x) \right], \\
 g_1^n (x) &=& \frac{1}{9} \,\left[ \,\Delta u(x) + 
 4 \,\Delta d(x) - \Delta s(x) \right].
\end{eqnarray}
Assuming that the polarizations of the $s$- and $\bar{s}$-quarks
are small, we therefore have an approximate relation
\begin{equation}
 g_1^N (x) \ \simeq \ \frac{5}{36} \,\Delta \Sigma (x),
\end{equation}
with
\begin{equation}
 \Delta \Sigma (x) \ \simeq \ 
 ( \Delta u(x) + \Delta \bar{u}(x) ) \ + \ 
 ( \Delta d(x) + \Delta \bar{d}(x) ) ,
\end{equation}
which denotes that $g_1^N (x)$ is proportional to the isosinglet
quark helicity distribution,
the integral of which gives the intrinsic quark-spin contribution to
the nucleon spin sum rule. This is of course exact only at the
leading-order QCD and under the assumption of small strange quark
polarization. Still, it clearly indicates the importance of precise
measurements of spin-dependent structure function of the deuteron
$g_1^d (x)$, which has recently been carried out by the COMPASS
group \cite{COMPASS06D} and also by the HERMES
group \cite{HERMES06D}.

Before comparing the predictions of the CQSM with the new COMPASS data
for $g_1^d (x)$, several remarks are in order.
Our predictions are based on the longitudinally polarized quark
distributions evaluated in \cite{WK99} within the framework of
flavor $SU(2)$ CQSM and those evaluated in \cite{Waka03}
within the framework of flavor $SU(3)$ CQSM.
The $SU(2)$ CQSM is essentially {\it parameter free},
since its only one model parameter, i.e. the dynamical quark mass
$M$ \cite{DPP88} was already fixed to be
$M \simeq 375 \,\mbox{MeV}$ from the
analyses of low energy nucleon observables. On the other hand,
the $SU(3)$ CQSM contains one additional parameter, i.e. the
mass difference $\Delta m_s$ between the strange and nonstrange
quarks. In \cite{Waka03}, the value of $\Delta m_s$ was fixed to
be $\Delta m_s \simeq 100 \,\mbox{MeV}$ so as to reproduce the
empirical unpolarized distribution of strange quarks.
In the case of $SU(2)$ model, we regard the theoretical quark
distributions $\Delta u(x), \Delta d(x), \Delta \bar{u}(x)$,
and $\Delta \bar{d}(x)$ as initial parton distributions
prepared at the low energy model scale around $Q_{ini}^2 = 0.3 \,
\mbox{GeV}^2 \simeq (600 \,\mbox{MeV})^2$, following the spirit
of the QCD analysis by Gl\"{u}ck, Reya and Vogt \cite{GRV95}.
The polarized strange quark distributions
$\Delta s(x)$ and $\Delta \bar{s} (x)$
as well as the polarized gluon distribution $\Delta g(x)$ are all set
zero at this low energy scale. We then solve the standard
DGLAP equation at the NLO to obtain the parton distributions
and the relevant structure functions at the high energy scale.
In the case of $SU(3)$ model, only a difference is that it can provide
us with the theoretical polarized strange quark distributions
$\Delta s(x)$ and $\Delta \bar{s}(x)$ as well at the initial model
energy scale.

\begin{figure}[htb]
\begin{center}
  \includegraphics[height=.40\textheight]{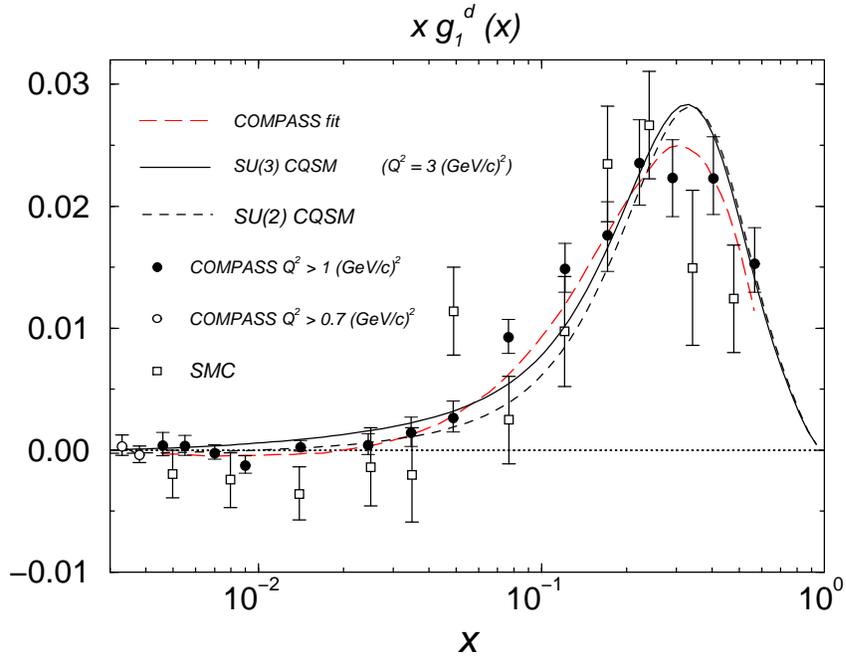}
  \caption{\baselineskip16pt The predictions of the $SU(2)$ and
  $SU(3)$ CQSM in comparison with the new COMPASS data for
  $x \,g_1^d (x)$   (filled circles) and their NLO QCD fits
  (long-dashed curve).
  The two COMPASS points at low $x$ (low $Q^2$), which are not
  included in their QCD fits, are also shown by open circles.
  Here, the theoretical predictions correspond to the fixed energy
  scale $Q^2 = 3 \,(\mbox{GeV/c})^2$, which corresponds to
  the average $Q^2$ of the COMPASS data, while the COMPASS points
  are given at the $\langle Q^2 \rangle$ where they were measured.
  The old SMC data \cite{SMC98} transformed to the corresponding
  COMPASS points are also shown by open squares, for reference.}%
\label{fig1:xg1d}
\end{center}
\end{figure}

Fig.\ref{fig1:xg1d} shows the comparison between our predictions for
$x \,g_1^d (x,Q^2)$ given several years ago and the new COMPASS data
(filled and open circles) together with the old SMC data
(open squares). 
The solid and the dashed curves respectively stand for the predictions
of the $SU(3)$ and $SU(2)$ CQSM evolved to the energy scale
$Q^2 = 3 \,\mbox{GeV}^2$, which is the average energy scale of the
new COMPASS measurement. The long-dashed curve shown for reference
is the next-to-leading order QCD fit by the COMPASS
group \cite{COMPASS06D}.
As one can see, the new COMPASS data show a considerable deviation
from the central values of the old SMC data in the small $x$ region.
One finds that our predictions are consistent with the new
COMPASS data especially in the small $x$ region.

\begin{figure}[bht]
\begin{center}
  \includegraphics[height=.40\textheight]{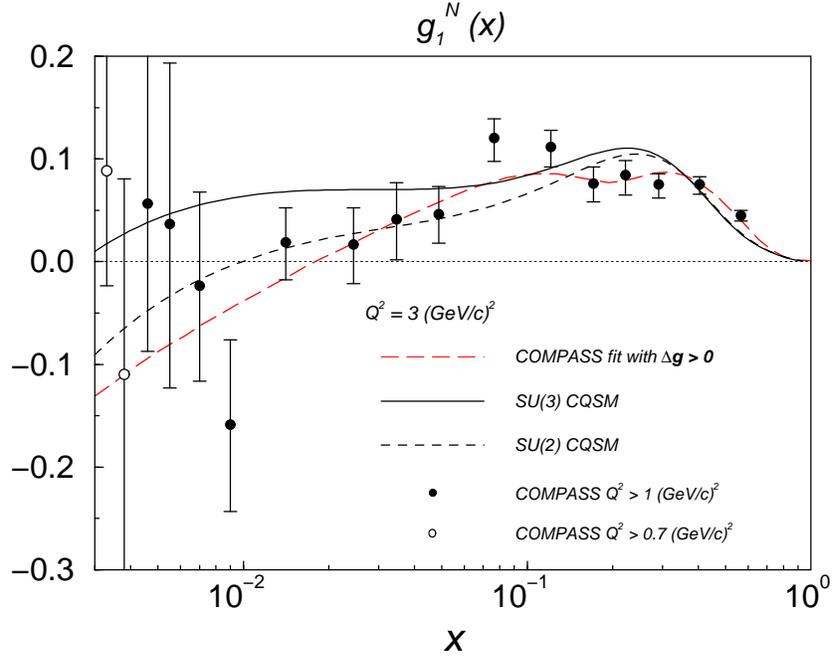}
  \caption{\baselineskip16pt The predictions of the $SU(2)$ and
  $SU(3)$ CQSM in comparison with the new COMPASS data for $g_1^N (x)$
  (filled and open circles) and their NLO QCD fits (long-dashed curve).}%
\label{fig2:g1N}
\end{center}
\end{figure}

This tendency can more clearly be seen in the comparison of
$g_1^N (x)$ illustrated in Fig.\ref{fig2:g1N}.
The filled circles here represent the new
COMPASS data for $g_1^N (x)$ evolved to the common energy scale
$Q^2 = 3 \,(\mbox{GeV/c})^2$, while the long-dashed curve is the
result of the next-to-leading order QCD fit by the COMPASS group
at the same energy scale. The corresponding
predictions of the $SU(3)$ and $SU(2)$ CQSM are represented by the
solid and dashed curves, respectively. For the quantity $g_1^N (x)$,
the experimental uncertainties are still fairly large in
the small $x$ region. Still, one can say that the predictions of
the CQSM is qualitatively consistent with the new COMPASS data
as well as their QCD fit.

\begin{figure}[htb]
\begin{center}
  \includegraphics[height=.4\textheight]{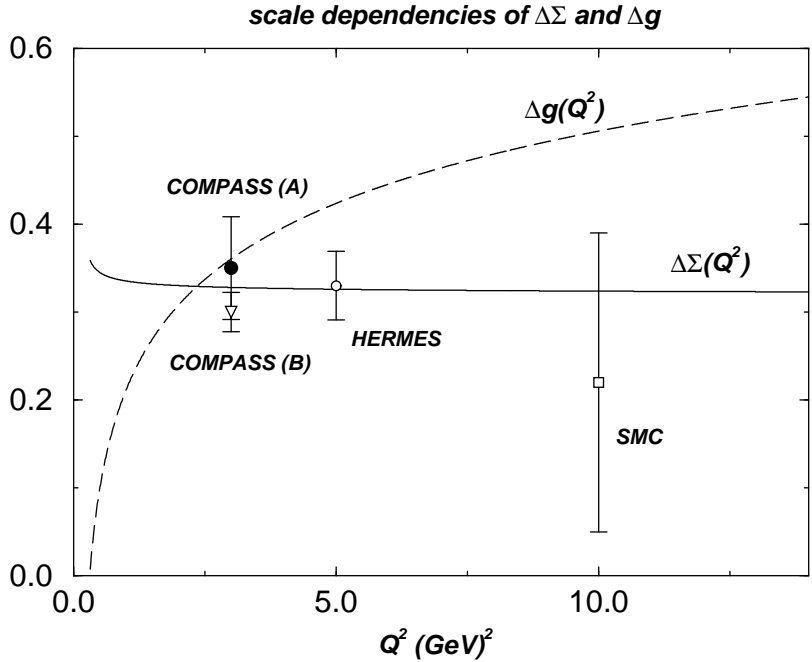}
  \caption{\baselineskip16pt The scale dependencies of
  $\Delta \Sigma$ and $\Delta g$
  predicted by the CQSM in combination with the NLO DGLAP equation
  are compared with the recent QCD fits by the COMPASS group 
  (filled circle and open triangle) and by the
  HERMES group (open circle). The old SMC result is also
  shown by an open square.}
\label{fig3:evolsig}
\end{center}
\end{figure}

The COMPASS group also extracted the matrix element of the
flavor-singlet axial charge $a_0$ \cite{COMPASS06D},
which can be identified with
the net longitudinal quark polarization $\Delta \Sigma$ in the
$\overline{\rm MS}$ factorization scheme. Taking the value of
$a_8$ from the hyperon beta decay, under the assumption of
$SU(3)$ flavor symmetry, they extracted from the QCD fit of the
new COMPASS data for $g_1^d(x)$ the value of $\Delta \Sigma$ as
\begin{equation}
 \Delta \Sigma (Q^2 = 3 \,\mbox{GeV}^2)_{COMPASS \,(A)} \ = \ 0.35 
 \ \pm \ 0.03 \,(stat.) \ \pm \ 0.05 \,(syst.) .
\end{equation}
On the other hand, the same quantity derived from the fits to all
$g_1$ data is a little smaller
\begin{equation}
 \Delta \Sigma (Q^2 = 3 \,\mbox{GeV}^2)_{COMPASS \,(B)} \ = \ 0.30
 \ \pm \ 0.01 \,(stat.) \ \pm \ 0.02 \,(evol.) .
\end{equation}
A similar analysis was also reported by the HERMES
group \cite{HERMES06D}. Their result is
\begin{equation}
 \Delta \Sigma (Q^2 = 5 \,\mbox{GeV}^2)_{HERMESS} \ = \ 0.330
 \ \pm \ 0.011 \,(theor.) \ \pm \ 0.025 \,(exp.) \ \pm \ 0.028 \,(evol.) .
\end{equation}

Main changes of these new QCD analyses from the old SMC analysis
\cite{SMC98} are considerable reduction of error bars and
upward shift of the central values.
Moreover, the results of the two groups for
$\Delta \Sigma$ look mutually consistent within the reduced error bars.
We now compare these new results with
the prediction of the $SU(3)$ CQSM given in our previous papers.
Shown in Fig.\ref{fig3:evolsig} are the prediction of the CQSM
for $\Delta \Sigma$
and $\Delta g$ as functions of the energy scale $Q^2$.
They are obtained
by solving the standard DGLAP equation at the NLO with the
prediction of the model as the initial condition given at the scale
$Q_{ini}^2 = 0.30 \,\mbox{GeV}^2 \simeq (600 \,\mbox{MeV})^2$.
Since the CQSM is an effective quark model, which contains no
gluon degrees of freedom, $\Delta g$ is simply assumed to be zero
at the initial scale. One sees that the new COMPASS and the
HERMES results for $\Delta \Sigma$ are surprisingly close to
the prediction of the CQSM. Also interesting is the longitudinal
gluon polarization $\Delta g$. In spite that we have assumed
that $\Delta g$ is zero at the starting energy, it grows rapidly
with increasing $Q^2$. As pointed out in \cite{Cheng96},
the growth of the gluon polarization with $Q^2$ can be traced
back to the positive
sign of the anomalous dimension $\gamma^{(0)1}_{qg}$. The positivity
of this quantity dictates that the polarized quark is preferred to
radiate a gluon with helicity parallel to the quark polarization.
Since the net quark spin component in the proton is positive,
it follows that $\Delta g > 0$ at least for the gluon perturbatively
emitted from quarks. The growth rate of $\Delta g$ is so fast
especially in the relatively small $Q^2$ region that its magnitude
reaches around $(0.3 - 0.4)$ already at $Q^2 = 3 \,\mbox{GeV}^2$, which
may be compared with the estimate given by the COMPASS group
\cite{COMPASS06D} :
\begin{equation}
 \Delta g (Q^2 = 3 \,\mbox{GeV}^2)_{COMPASS} \ \simeq \ 
 (0.2 - 0.3).
\end{equation}

\begin{figure}[htb]
\begin{center}
  \includegraphics[height=.4\textheight]{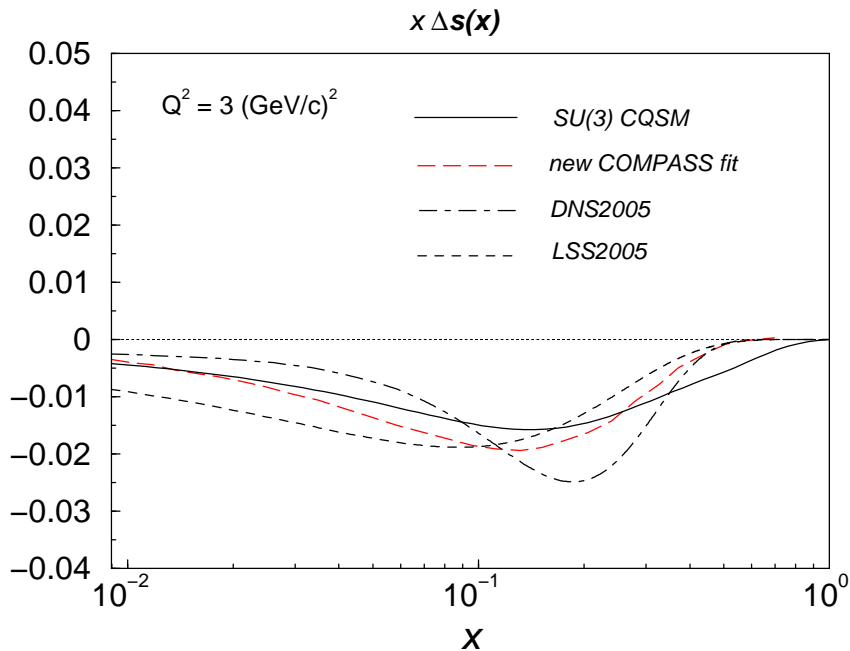}
  \caption{\baselineskip16pt The prediction of the $SU(3)$ CQSM
  for the polarized strange quark distribution
  $x \,\Delta s(x)$
  is compared with the recent QCD fits by the COMPASS group
  (long-dashed curve). The corresponding distributions from the
  DNS2005 and the LSS2005 fits are also shown for comparison by the
  dash-dotted and dashed curves, respectively.}
\label{fig4:xds}
\end{center}
\end{figure}

Also interesting to investigate is the COMPASS fits for the polarized
strange quark distributions,
extracted from the difference between
$\Delta \Sigma (x)$ and $\Delta q_8 (x)$. They performed two
next-to-leading order fits corresponding to positive and negative
gluon polarizations.
The long-dashed curve in Fig.\ref{fig4:xds} shows the
polarized strange quark distribution
$x \,\Delta s(x)$ at $Q^2 = 3 \,\mbox{GeV}^2$
corresponding to the fits with $\Delta g > 0$, while the
solid curve represents the corresponding predictions of the $SU(3)$
CQSM. For comparison, we also show the corresponding distributions
from the DNS2005 \cite{DNS2005} and LSS2005 \cite{LSS2005} QCD
fits. Note that, the flavor symmetry of the polarized strange sea,
i.e. $\Delta s(x) = \Delta \bar{s}(x)$ is assumed in all the above
three QCD fits. On the other hand, within the CQSM, as was pointed
out in \cite{Waka03}, the longitudinal strange quark polarization
is almost solely born by the $s$-quark and the polarization of
$\bar{s}$-quark is very small,
Bearing this fact in mind, one sees that the result of
the new COMPASS fits for $x \,\Delta s(x)$
is definitely negative and its magnitude is qualitatively consistent
with the prediction of the CQSM as well as with the DNS2005 and LSS2005
QCD fits.

The net strange quark polarization
$\Delta s + \Delta \bar{s}$, or the first moment of
$\Delta s(x) + \Delta \bar{s}(x)$ extracted by the COMPASS and the
HERMES group may also be interesting to see. The COMPASS group
obtained
\begin{equation}
 (\Delta s + \Delta \bar{s}) (Q^2 \rightarrow \infty)_{COMPASS}
 \ = \ - \,0.08 \ \pm 0.01 \,(stat.) \ \pm \ 0.02 \,(stat.),
\end{equation}
while the result of the HERMES analysis is
\begin{equation}
 (\Delta s + \Delta \bar{s}) (Q^2 = 5 \,\mbox{GeV}^2)_{HERMES}
 \ = \ - \,0.085 \ \pm 0.013 \,(theor.) \ \pm \ 0.008 \,(exp.)
 \ \pm \ 0.028 \,(evol.) .
\end{equation}
One finds that the results of the two semi-empirical fits are not
only mutually consistent but also they are surprisingly close to the
the corresponding prediction of the $SU(3)$ CQSM given by
\begin{equation}
 (\Delta s + \Delta \bar{s}) (Q^2 = 5 \,\mbox{GeV}^2)_{CQSM}
 \ = \ - \,0.082.
\end{equation}

A sizable polarization of the strange sea appears to contradict
the indication of the semi-inclusive DIS analysis \cite{HERMES99}.
However, we believe that our understanding of the semi-inclusive
processes has not reached the precision of inclusive DIS physics yet.
We also emphasize that the large and negative polarization of
the strange quarks is not a crucial factor for our resolution
scenario of the nucleon spin puzzle.
This is clear from the fact that 
the flavor SU(2) CQSM, which naturally predicts zero
strange polarization at the model energy scale, already explains
small $\Delta \Sigma$. In fact, aside from very small SU(3) breaking
effect, which turns out to be the order of 0.01, both of the SU(2)
CQSM and the SU(3) CQSM gives exactly the
same answer for $\Delta \Sigma$ as
\begin{eqnarray}
 \Delta \Sigma [SU(2)] &\equiv& \Delta u + \Delta \bar{u} + 
 \Delta d + \Delta \bar{d} \ = \ 0.35, \\
 \Delta \Sigma [SU(3)] &\equiv& \Delta u + \Delta \bar{u} + 
 \Delta d + \Delta \bar{d} + \Delta s + \Delta \bar{d} \ = \ 0.35.
\end{eqnarray}
Since $\Delta s + \Delta \bar{s} < 0$ in the SU(3) CQSM, this means that
$\Delta u + \Delta \bar{u} + \Delta d + \Delta \bar{d}$ in the SU(2) model
is smaller than that in the SU(3) CQSM, while keeping the equality
\begin{equation}
 \Delta \Sigma [SU(2)] \ = \ \Delta \Sigma [SU(3)].
\end{equation}
In any case, the above explanation clearly shows that the unique
feature of the CQSM, which can reproduce very small $\Delta \Sigma$,
is not crucially dependent on the negative polarization of strange
quarks, but it rather comes from the basic dynamical assumption of
the model, i.e. the physical picture of the nucleon as a rotating
hedgehog, which naturally generates large quark orbital angular momentum.
As a consequence, the HERMES result, even though it
is assumed to be correct, would not change the main conclusions
of the present paper, i.e. the resolution scenario of the nucleon
spin puzzle based on the importance of the quark orbital
angular momentum. For other resolution scenarios of the nucleon spin
puzzle, we refer to the recent workshop summary \cite{BA06}.

To conclude, the new measurements of the deuteron spin-structure
function $g_1^d (x)$ carried out by the COMPASS group as well as
by the HERMES group achieved a remarkable improvement in the
accuracy of the experimental data, especially in the low
$x$ region, as compared with the existing old data.
As an important outcome, our knowledge on
the net quark helicity contribution $\Delta \Sigma$ to the total nucleon
spin has been improved to a large degree.
As we have pointed out, the value of $\Delta \Sigma$ extracted from
the new QCD fits by the COMPASS and the HERMES groups is around
$0.3 \sim 0.35$, which is surprisingly close to the prediction of
the CQSM. Now that the role of quark helicity contribution to the
nucleon spin sum rule has been understood fairly well, we come back to
the question : what carry the rest of the nucleon spin?
The CQSM claims that the role of quark orbital momentum
is important at least at the low energy scale of nonperturbative QCD
around $Q^2 \simeq (600 \,\mbox{MeV})^2$.
(Although this is a highly model-dependent statement, we can give a
kind of model-independent analysis, based only upon some reasonable
theoretical postulates, which supports the importance of quark OAM
at the low energy scale \cite{Waka05},\cite{WN06}.)
We hope that this unique prediction of the CQSM will be verified
by the near-future measurement of the generalized parton
distribution functions of the nucleon with enough precision.

\vspace{10mm}
%\newpage
\noindent
\begin{large}
{\bf Acknowledgement}
\end{large}

\vspace{3mm}
This work is supported in part by a Grant-in-Aid for Scientific
Research for Ministry of Education, Culture, Sports, Science
and Technology, Japan (No.~C-16540253)

%
%  Reference
%

\setlength{\baselineskip}{5mm}


\begin{thebibliography}{99}

\bibitem{EMC88}
EMC Collaboration, J.~Ashman~et al.,
Phys. Lett. B206 (1988) 364 ;\\
Nucl. Phys. B328 (1989) 1.

\bibitem{COMPASS06G}
COMPASS Collaboration, E.S.~Ageev~et al.,
Phys. Lett. B633 (2006) 25.

\bibitem{PHENIX06}
PHENIX Collaboration, K.~Boyle~et al.,
nucl-ex/0606008.

\bibitem{STAR05}
STAR Collaboration, J.~Kiryluk~et al.,
hep-ex/0512040.

\bibitem{STAR06}
STAR Collaboration, R.~Fatemi~et al.,
nucl-ex/0606007.

\bibitem{BG06}
S.J.~Brodsky and S.~Gardner,
hep-ph/0608219.

\bibitem{Sivers90}
D.W.~Sivers,
Phys. Rev. D41 (1990) 83 ;
Phys. Rev. D43 (1991) 261.

\bibitem{Burk02}
M.~Burkardt,
Phys. Rev. D66 (2002) 114005.

\bibitem{Ji98}
X.~Ji,
J. Phys. G24 (1998) 1181.

\bibitem{BEK88}
S.J.~Brodsky, J.~Ellis, and M.~Karliner,
Phys. Lett. B206 (1988) 309.

\bibitem{WY91}
M.~Wakamatsu and H.~Yoshiki,
Nucl. Phys. A524 (1991) 561.

\bibitem{WK99}
M.~Wakamatsu and T.~Kubota,
Phys. Rev. D60 (1999) 034020.

\bibitem{Waka03}
M.~Wakamatsu,
Phys. Rev. D67 (2003) 034005 ;
Phys. Rev. D67 (2003) 034006.

\bibitem{COMPASS05D}
COMPASS Collaboration, E.S.~Ageev~et al.,
Phys. Lett. B612 (2005) 154.

\bibitem{COMPASS06D}
COMPASS Collaboration, V.Yu.~Alexakhin~et al.,
hep-ex/0609038.

\bibitem{HERMES06D}
HERMES Collaboration, A.~Airapetian~et al.,
hep-ex/0609039.

\bibitem{GRV95}
M.~Gl\"{u}ck, E.~Reya, and A.~Vogt,
Z. Phys. C67 (1995) 433.

\bibitem{DPP88}
D.I.~Diakonov, V.Yu.~Petrov, and P.V.~Pobylitsa,
Nucl. Phys. B306 (1988) 809.

\bibitem{SMC98}
SMC Collaboration, B.~Adeva~et al.,
Phys. Rev. D58 (1998) 112001.

\bibitem{Cheng96}
H.-Y.~Cheng,
Int. J. Mod. Phys. A11 (1996) 5109.

\bibitem{DNS2005}
D.de~Florian, G.A.~Navarro, and R.~Sassot,
Phys. Rev. D71 (2005) 094018.

\bibitem{LSS2005}
E.~Leader, A.V.~Sidorov, and D.B.~Stamenov,
Phys. Rev. D73 (2006) 034023.

\bibitem{HERMES99}
HERMES Collaboration, A.~Ackerstaff~et al.,
Phys. Lett. B464 (1995) 123.

\bibitem{BA06}
S.D.~Bass and C.A.~Aidala,
hep-ph/0606269.

\bibitem{Waka05}
M.~Wakamatsu,
Phys. Rev. D72 (2005) 074006.

\bibitem{WN06}
M.~Wakamatsu and Y.~Nakakoji,
Phys. Rev. D74 (2006) 054006.

\end{thebibliography}
\end{document}